\documentclass[a4paper]{jpconf}
\usepackage{graphicx}

\begin{document}

\newcommand{\bb}{\mbox{$\beta\beta$}}
\newcommand{\nue}{\mbox{$\nu_e$}}
\newcommand{\ise}{\mbox{$^{82}$Se}}
\newcommand{\um}{\mbox{$\mu$m}}
\newcommand{\Tm}{\mbox{\emph{Topmetal II$^{\rm -}$}}}

\title{The Selena Neutrino Experiment}

\author{A.E.~Chavarria$^1$ for the Selena Collaboration}
\address{$^1$Center for Experimental Nuclear Physics and Astrophysics, University of Washington, Seattle, WA 98195, United States}

\ead{chavarri@uw.edu}

\begin{abstract}
Imaging sensors made from an ionization target layer of amorphous selenium (aSe) coupled to a silicon complementary metal-oxide-semiconductor (CMOS) active pixel array for charge readout are a promising technology for neutrino physics. The high spatial resolution in a solid-state target provides unparalleled rejection of backgrounds from natural radioactivity in the search for neutrinoless \bb\ decay and for solar neutrino spectroscopy with \ise . We present results on the ionization response of aSe measured from the photoabsorption of 122\,keV $\gamma$ rays in a single-pixel device. We report on the progress in the fabrication and testing of the first prototype imaging sensors based on the \Tm\ pixelated CMOS charge readout chip. We explore the scientific reach of a large neutrino detector with the proposed technology based on our experimental understanding of the sensor performance.
\end{abstract}

\section{Overview}

\begin{figure}[t]
	\begin{center}
		\includegraphics[width=\textwidth]{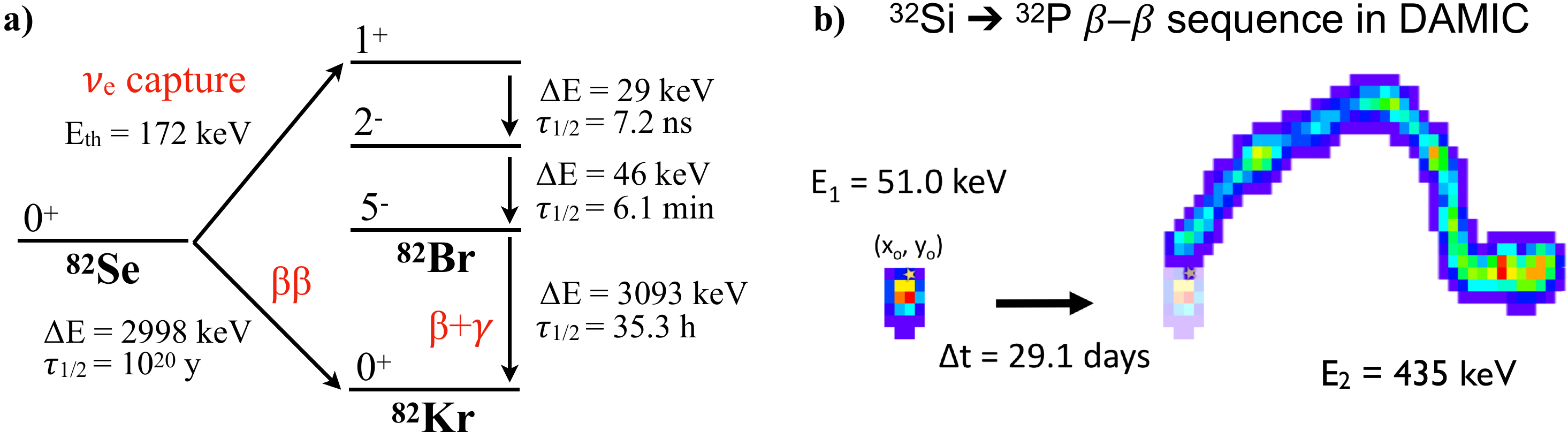}
		\caption{{\bf a)} Diagram showing two natural channels for $^{82}$Se to transmute into $^{82}$Kr. Both $\beta\beta$ decay and solar $\nu_e$ capture will be detected with high efficiency by Selena for nuclear physics studies. {\bf b)} The identification of a single $^{32}$Si atom by DAMIC at SNOLAB: a sequence of two $\beta$ decays separated in time by many days and with the start points of the electron tracks detected at the same location. Pixels toward the red end of the spectrum collect more charge.}
		\label{fig:selena_principle}
	\end{center}
\end{figure}

The Selena experiment was proposed in Ref.~\cite{Chavarria:2016hxk} to search for the neutrinoless $\beta\beta$ decay of $^{82}$Se with unprecedented sensitivity.
Recently, we have expanded its scientific case to include solar neutrino spectroscopy based on $\nu_e$ capture on $^{82}$Se.
Figure~\ref{fig:selena_principle}a shows the two nuclear reactions of interest for Selena.

The Selena detector will consist of towers of imaging modules made from $\sim$1\,mm-thick amorphous selenium (aSe)\textemdash isotopically enriched in $^{82}$Se\textemdash deposited on a CMOS active pixel charge sensor (APS).
Free charge generated by ionizing particles in the aSe is drifted by an applied electric field and collected by the pixels resulting in high resolution images.
From these images, Selena can $\emph{i)}$~measure the deposited energy, $\emph{ii)}$~ identify the type and number of ionizing particles (electrons, $\alpha$'s, etc.) from the track topologies,  and $\emph{iii)}$~identify radioactive decay sequences by spatio-temporal correlations.
This strategy has been realized to some extent in the context of dark matter searches by the DAMIC experiment to constrain the activities in the silicon target of $^{32}$Si, and every isotope in the $^{238}$U and $^{232}$Th decay chains~\cite{DAMIC:2020wkw}.
For example, Fig.~\ref{fig:selena_principle}b shows the identification of a $^{32}$Si decay ($Q$-value $=225$\,keV, $\tau_{1/2}=150$\,y) from its spatial correlation with the $\beta$ track of its daughter $^{32}$P ($Q$-value $=1.71$\,MeV, $\tau_{1/2}=14$\,d) many days later.
Note that start and end points of the $\beta$ track from $^{32}$P decay can be distinguished by the presence of a high density of charge, \emph{i.e.}, the Bragg peak, at the end of the track.
In the context of Selena, this experimental strategy will allow for background-free spectroscopy of $\beta\beta$ decay and solar neutrinos in exposures $>$100\,ton-year.

Selena is particularly efficient in its use of the isotopically-enriched active target since background suppression relies on event-by-event identification of the signal and not self shielding.
Furthermore, our background estimates assume background event rates comparable to those already achieved by kg-scale solid-state detectors for rare event searches, \emph{e.g.}, the M{\footnotesize AJORANA} D{\footnotesize EMON\-STRAT\-OR}~\cite{Majorana:2019nbd}.
Thus, the requirements in the radiopurity of the construction materials and processes, although challenging, are already possible.

Finally, we note that Selena will operate at room temperature.
The CMOS APS for its imaging modules will be fabricated with standard CMOS foundry processes, and the aSe deposited following industrial-scale processes developed for the fabrication of medical devices.
A 10\,ton Selena detector would consist of 50,000 imaging modules, each with 200\,g of amorphous $^{82}$Se sandwiched between two large area CMOS APS fabricated on 300-mm diameter wafers.

\section{R\&D status}

Our initial experimental efforts succeeded in \emph{i)}~measuring the energy response of aSe, and \emph{ii)}~demonstrating single-electron imaging in a hybrid aSe/CMOS sensor.

\begin{figure}[t]
	\begin{center}
		\includegraphics[width=\textwidth]{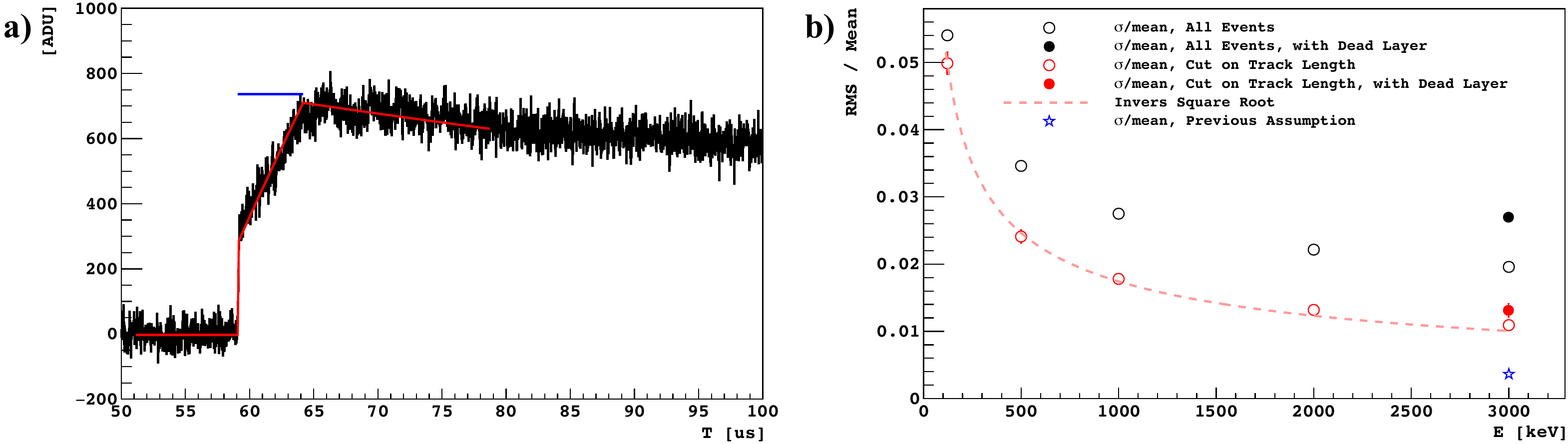}
		\caption{{\bf a)} Ionization signal pulse from a 122\,keV $\gamma$ ray absorbed in an aSe single-pixel sensor operated at a drift field of 30\,V/$\mu$m. The red line shows the best-fit to the signal pulse. The horizontal blue line marks the fitted pulse height. {\bf b)} Extrapolation by simulation of the energy resolution in aSe at 50\,V/$\mu$m based on the measured response to 122 keV $\gamma$ rays. The black markers show the resolution if the energy is estimated as the total ionization charge. The red markers show the improvement obtained after scaling linearly by the number of pixels in the track ($\sim$track length). The filled markers show the effect of a 10\,$\mu$m-thick dead layer between imaging modules. The blue star is the best-case expectation from Ref.~\cite{Chavarria:2016hxk}.}
		\label{fig:selena_ionization}
	\end{center}
\end{figure}

The energy resolution at $Q_{\beta\beta}$ determines the capability of Selena to spectrally resolve the neutrinoless $\beta\beta$ decay signal over the two-neutrino $\beta\beta$ decay background spectrum.
In Ref.~\cite{Li:2020ryk}, we reported our measurement of the response of aSe to 122\,keV $\gamma$ rays from a $^{57}$Co radioactive source.
The measurement was performed with a ``single-pixel'' sensor: a rectangular 200\,$\mu$m-thick layer of aSe (of natural isotopic abundances) sandwiched between two electrodes, with the anode connected to a high voltage (HV) and the cathode to a low-capacitance CMOS pulse-reset charge-sensitive amplifier.
The aSe deposition was performed by our collaborators at Hologic Corporation, who use aSe as the target for their commercial flat-panel X ray detectors for radiography~\cite{rowlands}.
Ionization charge generated from the photoabsorption of 122\,keV $\gamma$ rays in aSe is drifted toward the electrodes and generates signal pulses as shown in Figure~\ref{fig:selena_ionization}a.
Since holes have a much higher mobility than electrons in aSe, pulses closer to the anode (cathode) have faster (slower) rise times.
This sensor features the lowest noise ever achieved in aSe, which allowed us to not only measure precisely the ionization signal amplitude but also to reconstruct the depth of the interaction from the pulse shape for the first time.

We acquired data as a function of HV and measured the total ionization charge yield and the fractional energy resolution at 122\,keV.
We reproduced the data with a full detector simulation, which started with particle tracking with {\tt Geant4} followed by a recombination model, charge-carrier transport and a custom electronics simulation.
We conclude that there must be a strong dependence of recombination on the ionization density along the electron tracks, \emph{i.e.}, the charge yield along the electron tracks has an inverse relation with $dE/dx$.
Thus, if the energy of the event is estimated from the total charge, the energy resolution is dominated by fluctuations in $dE/dx$ and not charge-carrier statistics.
Figure~\ref{fig:selena_ionization}b shows the extrapolation of our energy resolution by simulation to $Q_{\beta\beta}$.
A simple linear correction to the event energy based on the length of the imaged electron tracks in a simulated Selena detector shows a significant improvement in the energy resolution to 1.2\% RMS. Although promising, this is still above the 0.4\% limit from carrier statistics ($1/\sqrt{N_{\mbox{\scriptsize e-h}}}$).

\begin{figure}[t]
	\begin{center}
		\includegraphics[width=\textwidth]{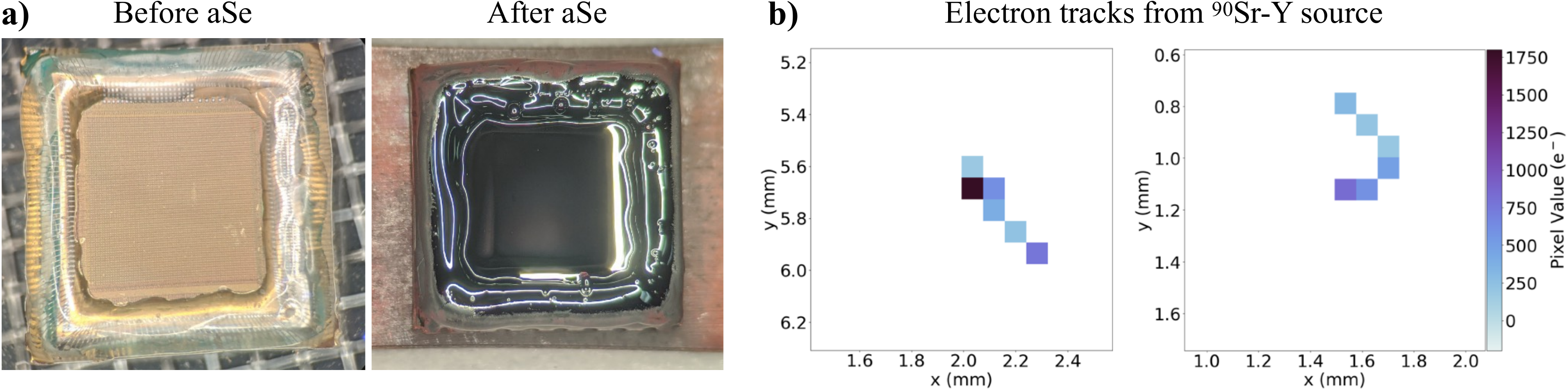}
		\caption{{\bf a)} \emph{Topmetal-II$^{\mbox{-}}$} CMOS APS before and after aSe deposition. {\bf b)} Electron tracks observed with our first hybrid aSe/CMOS sensor operated at 4\,V/$\mu$m at room temperature.}
		\label{fig:selena_topmetal}
	\end{center}
\end{figure}

Recently, we successfully coupled an aSe target layer to a CMOS APS, with which we achieved the lowest pixel noise ever in an aSe imager of 23\,$e^-$ RMS. 
We started with the \emph{Topmetal-II$^{\mbox{-}}$} CMOS APS from Berkeley Lab~\cite{An:2015oba}, which features a rectangular array of 72$\times$72 pixels with 83\,$\mu$m pitch.
A 500\,$\mu$m thick layer of aSe was deposited by Hologic on top of the bare metallic contacts of the chip (Figure~\ref{fig:selena_topmetal}a), followed by a thin gold contact for the HV.
The sensor was operated in a test setup that includes mechanical components to mount the sensor, radioactive and light sources, and HV and front-end electronics.
The \emph{Topmetal-II$^{\mbox{-}}$} was driven in rolling shutter mode with an external clock generator, with the output of the pixel charge-sensitive amplifiers (CSA) multiplexed into a single channel that was sampled continuously.
An FPGA monitors the \emph{Topmetal-II$^{\mbox{-}}$} output and sends a trigger when any pixel value returns a signal above its baseline level, which causes the pixel array output to be digitized with high resolution for the six subsequent frames.
An offline reconstruction software converts the \emph{Topmetal-II$^{\mbox{-}}$} output into images.
Data were acquired with $^{57}$Co and $^{90}$Sr-Y sources (Figure~\ref{fig:selena_topmetal}b), which demonstrate single electron detection in a hybrid aSe/CMOS pixelated sensor for the first time.

\section{Neutrinoless \bb\ decay of \ise }

Our initial study on the expected backgrounds for a neutrinoless \bb\ decay search predicts background rates at $Q_{\beta\beta}=3$\,MeV below $6\times10^{-5}$ per keV per ton-year \cite{Chavarria:2016hxk}.
This extremely low background level is possible because of a combination of factors.
First, the high $Q_{\beta\beta}$ of \ise\ is at an energy greater than most backgrounds from primordial $^{238}$U and $^{232}$Th radiocontaminants, which leads to a relatively low raw event rate.
Second, spatio-temporal correlations effectively reject any radioactive decays in the bulk or the surfaces of the imaging modules.
Finally, external $\gamma$-ray backgrounds, which mostly produce single-electron events from Compton scattering or photoelectric absorption, are suppressed by the requirement that the \bb\ signal events have two clearly identified Bragg peaks (this selection retains 50\% of signal events while rejecting 99.9\% of the single-electron background).

The superb background suppression does not require much more than a realistic pixel pitch of 15\,$\mu$m and imager uptime close to 100\%, specifications that are within reach of current technology.
The spectroscopic identification of the neutrinoless \bb\ signal over the two-neutrino \bb\ decay background is a bigger challenge.
Figure~\ref{fig:selena_spectra}a shows the \bb\ decay spectrum assuming the energy resolution given by the red marker in Fig.~\ref{fig:selena_ionization}b, and $\tau_{1/2}=10^{28}$ for the neutrinoless channel.
Although there is significant interference from the two-neutrino channel, a limit of $\tau_{1/2}>10^{28}$\,y would be possible in a 100\,ton-year exposure.
However, to comfortably surpass the sensitivity of proposed next-generation experiments~\cite{Agostini:2017jim}, Selena must achieve an energy resolution closer to the 0.4\% limit from carrier statistics.
Fortunately, the energy reconstruction can be improved since there is significant information along the ionizing electron tracks that we are not currently exploiting.
We are exploring machine learning (ML) techniques to consider how $dE/dx$ evolves along the tracks and correct for the changing charge yield.

\section{Solar neutrino spectroscopy}

\begin{figure}[t]
	\begin{center}
		\includegraphics[width=\textwidth]{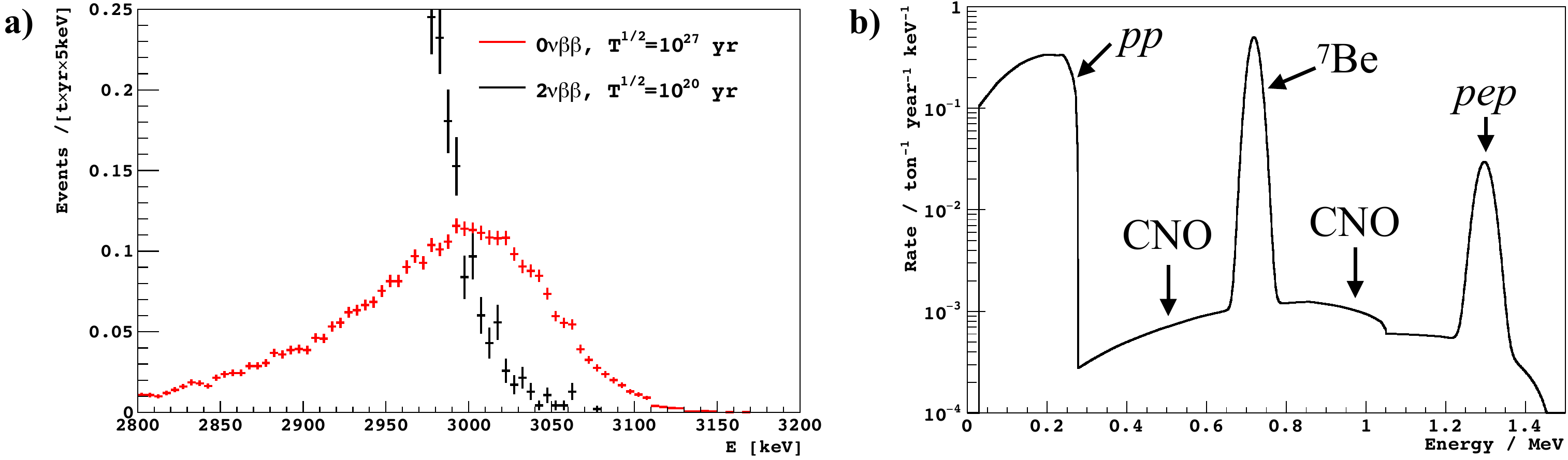}
		\caption{{\bf a)} Predicted spectrum from \bb\ decay of $^{82}$Se about $Q_{\beta\beta}$ in Selena, with $\tau_{1/2}$ for the zero- and two-neutrino decay channels given in the legend. {\bf b)} Spectrum of the prompt event from solar $\nu_e$ capture. The neutrino species from different solar-fusion reactions are labeled. Capture rates from Ref.~\cite{Frekers:2016haa}.}
		\label{fig:selena_spectra}
	\end{center}
\end{figure}

Solar $\nu_e$ captures in Selena can be tagged on an event-per-event basis with high efficiency to perform solar $\nu$ spectroscopy concurrently with the search for neutrinoless $\beta\beta$ decay.
The threshold for $\nu_e$ capture of 172\,keV provides sensitivity to all solar $\nu$ species, and leads to the decay sequence shown in Figure~\ref{fig:selena_principle}a.
The first step with $\tau_{1/2}=7.2$\,ns is too fast to be distinguished from the electron with energy $E_\nu-172$\,keV emitted following $\nu_e$ capture.
Together they constitute the ``prompt'' event, whose spectrum is shown in Figure~\ref{fig:selena_spectra}b with the energy resolution given by the red line in Fig.~\ref{fig:selena_ionization}b.
The prompt event is followed by a sequence of two decays, which are expected to occur at exactly the same spatial location with time delays of $\sim$5\,min and $\sim$1\,day, respectively.
Although these time separations may seem large, the probability of any two accidental events occurring at exactly the same pixel (\emph{i.e.}, in $\sim$10\,$\mu$g of aSe) is negligible.
Furthermore, any charged particle reaction to produce $^{82m}$Br  ($2^-$ state) from $^{82}$Se would have remarkably different prompt event topology.
Additionally, no background isotope has been identified that could mimic the $\nu_e$ capture sequence in time separation, event energies and topologies.
So far, it appears that zero-background solar $\nu$ spectroscopy with a large Selena detector may be possible.

\begin{table}[h]
	\begin{center}
		\begin{tabular}{lccc}
			Species &  $E$ range [keV] & $N$ in 100 ton-year & $1/\sqrt{N}$ \\
			\hline
			{\it pp} &	29--278	& 6170 & 1.3\% \\
			$^7$Be & 665--775  & 1850 & 2.3\% \\
			{\it pep} & 1230--1360 & 151 & 8.1\% \\
			CNO & 278--655, 785--1220& 63 & 13\% \\
			$^8$B & (1.5-15)$\times10^3$ & 209 & 6.9\% \\
		\end{tabular}
		
		\caption{Expected results for the capture rate of solar neutrinos from a counting experiment. $N$ is the expected number of $\nu_e$ captures for the different species in the specified energy ($E$) range. The last column provides an estimate of the statistical uncertainty in the measured rate.}
		\label{tab:selena_solar}
	\end{center}
\end{table}

Table~\ref{tab:selena_solar} presents the integrated count rates from Fig.~\ref{fig:selena_spectra}b in the specified energy regions.
With a 100\,ton-year exposure, Selena is expected to measure the $pp$ rate to $\sim$1\% for a strong constraint on the neutrino luminosity, the $pep$ rate to $\sim$8\% to probe the onset of the Mikheyev-Smirnov-Wolfenstein (MSW) effect for solar neutrinos~\cite{Mikheev:1987qk}, and the CNO-cycle rate to $\sim$10\% to measure solar metallicity~\cite{Serenelli:2009yc}.

The mean energy of the $^7$Be solar neutrino line is expected to be shifted relative to the neutrino line from $^7$Be electron-capture decay in the laboratory~\cite{Bahcall:1994cf}.
This energy shift of 1.3\,keV is directly correlated with the solar core temperature.
There is a corresponding shift of the \emph{pep} line of  7.6\,keV~\cite{carlos}.
We expect a Selena experiment with a 100\,ton-year exposure to measure these line shifts with high statistical significance for a $\sim$25\% estimate of the solar core temperature.

\section*{References}
\bibliographystyle{iopart-num}
\bibliography{myrefs}

\end{document}